\documentclass[pre,aps,12pt]{revtex4}
 \usepackage{graphicx}

\def\be{\begin{equation}}
\def\ee{\end{equation}}

\begin{document}

\title{Electron Clusters in Inert Gases}

\author{S.Nazin and V.Shikin \\
Institute of Solid State Physics of Russian Academy of Sciences \\
Chernogolovka, Moscow district, 142432, Russia}

\begin{abstract}
The paper addresses counterintuitive behavior of electrons
injected into dense cryogenic media with negative scattering
length $a_0$. Instead of expected polaronic effect (formation of
density enhancement clusters) which should substantially reduce
the electron mobility, an opposite picture is observed: with
increasing $|a_0|$ (the trend taking place for inert gases with
the growth of atomic number) and the medium density, the electrons
remain practically free. An explanation of this behaviour is
provided based on consistent accounting for the non-linearity of
electron interaction with the gaseous medium in the gas atom
number density.
\end{abstract}

\maketitle

PACS: 71.10.-w

One of the most interesting and still important issues in physics
of cryogenic media is the problem of electron clusters which
emerged almost simultaneously with that of electron bubbles.
However, it is much less transparent (compared to the case of
electron bubbles) from the experimental side. On the one hand,
there exist indications of the existence of electron clusters in
argon \cite{HuangFreeman}. On the other hand, they are not
observed on the expected scale in media with higher atomic
polarizabilities (krypton, xenon) which are presumably more likely
to develop various electron autolocalization phenomena. On the
contrary, the data on electron mobility in these media
\cite{MillerHoveSpear,Christophorou,LastBook} reveal that
electrons remain practically free (compared to mobility of
positive ions possessing the structure of massive polaronic-type
formations) in their motion, at least in the vicinity of the
characteristic electron mobility peak which is observed for all
heavy inert gases.

The existent description
\cite{Khrapak_ZHETF,Iakubov160,KhrapakBook} of electron clusters
in cryogenic media with negative scattering lengths $a_0$ employs
the well-known approximation \cite{Fermi1934,LifshitzSondheimer}
for electron-medium interaction energy which is linear in the gas
density $n$. Within this approximation, the minimal energy $V_0$
of delocalized electron injected into the gaseous media is
calculated as
    \be
         V_0=\frac{2\pi\hbar^2a_0}{m}n,
     \label{OpticalApprox}
    \ee
where $m$ is the free electron mass. In terms of electron energy
bands in solids, $V_0$ is the conduction band bottom energy. The
case of $a_0>0$ corresponds to formation a single-electron bubble.
On the other hand, a density enhancement domain with higher gas
atom concentration (i.e., a cluster) may develop around the
electron if $a_0<0$. The authors of Refs.
\cite{Khrapak_ZHETF,Iakubov160,KhrapakBook} made every effort to
provide a quantitatively accurate description of the gas density
around the localized electron in the linear approximation. In
addition to (\ref{OpticalApprox}), they also introduced a
non-local electron-gas interaction of the type
    \be
         V(r)=\int d^3r'v(r-r')\psi^2(r')
         \label{V_NonLocal}
    \ee
where $\psi(r)$ is the electron wave function, took into account
the deviation of the gas entropy contribution to the total free
energy from the ideal gas, etc. Their final conclusions
\cite{Khrapak_ZHETF,Iakubov160,KhrapakBook} practically coincide
with the intuitively expected picture: the electron cluster should
exist, and the electron localization degree as well as the cluster
mass should monotonously grow with the density media and
polarizability demonstrating exponential sensitivity to the
temperature. The outlined approach
\cite{Khrapak_ZHETF,Iakubov160,KhrapakBook} reveals no hints of
electron mobility growth with the medium density
\cite{MillerHoveSpear,Christophorou,LastBook}.

In the present paper we show that in gaseous media with negative
values of $a_0$ it is possible for electron, in a ceratin range of
gas densities and temperatures, to form an autolocalized state
involving formation of a cluster with the characteristic length
$\lambda\gg a_B$ (where $a_B$ is the Bohr radius) if the
electron-gas interaction $V_0(n)$ is treated beyond the linear
approximation in $n$ (\ref{OpticalApprox}). The paper is organized
as follows. First, the formal grounds for considering the
non-linear behaviour of $V_0(n)$. Then the electron cluster
structure is calculated within the non-linear approach. Finally,
the nature of deviations from non-linearity at small $n$ is
discussed.

\begin{figure}
\begin{center}
\includegraphics*[width=11cm]{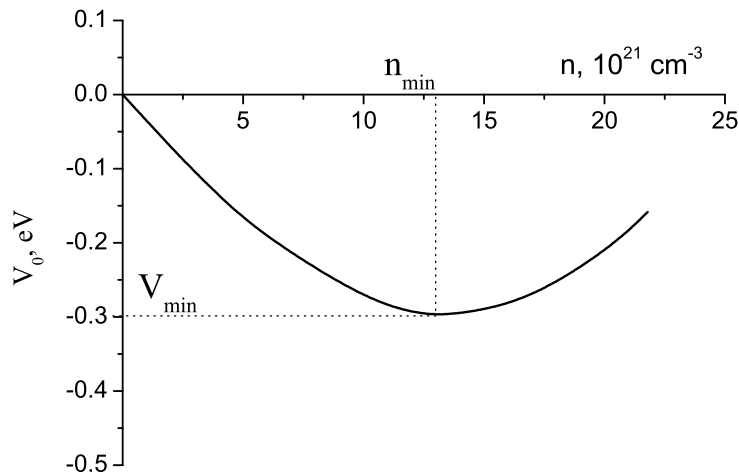}
\end{center}
\caption{Typical example of the minimal energy of delocalized
electron injected in heavy inert gases with negative scattering
length as a function of gas atom number density.}
\label{Fig_V_NonLin}
\end{figure}

As already mentioned earlier, the existent theory of autolocalized
electrons in cryogenic media employs the electron-gas interaction
(\ref{OpticalApprox}). However, the true minimal energy of an
electron injected in inert gases with negative scattering length
(Ar, Kr, Xe) is substantially non-linear. Direct experiments
\cite{NonLin_Experimental} reveal that the energy $V_0(n)$ is only
approximately linear at very small $n$ following
 $V_{min}$ at  a certain $n_{min}$ after which it grows again (see Fig. \ref{Fig_V_NonLin};
for Ar $n_{min}=13\cdot10^{21}$ cm$^{-3}$, $V_{min}=-0.3$ eV, for
$n_{min}=14\cdot10^{21}$ cm$^{-3}$, $V_{min}=-0.66$ eV, for
$n_{min}=11\cdot10^{21}$ cm$^{-3}$, $V_{min}=-0.83$ eV)). Most
important in our problem of cluster formation is the range of $n$
near $n_{min}$ where the derivative $\partial V_0/\partial n$
changes its sign. Indeed, it is natural to assume that dominating
in the problem of self-consistent calculation of the gas atom
number density $n(r)$ will be the densities minimizing the
electron-gas interaction energy. Hopefully, the quantitative
analysis can be based on any reasonable interpolation of the true
$V_0(n)$ reproducing the correct minimum depth and position. In
fact, we used the simplest polynomial approximation yielding in
addition the correct slope of $V_0(n)$ at small $n$:
    \be
        V_0(n) = \frac{2\pi \hbar^2 a_0 }{m}n(1 + An + Bn^2).
        \label{VFit}
    \ee
The parameters $A$ and $B$ were chosen for each inert gas in such
a way that the correct values \cite{NonLin_Experimental} of the
minimum depth $V_{min}$ and position $n_{min}$ were reproduced.
\begin{figure}
\begin{center}
\includegraphics*[width=11cm]{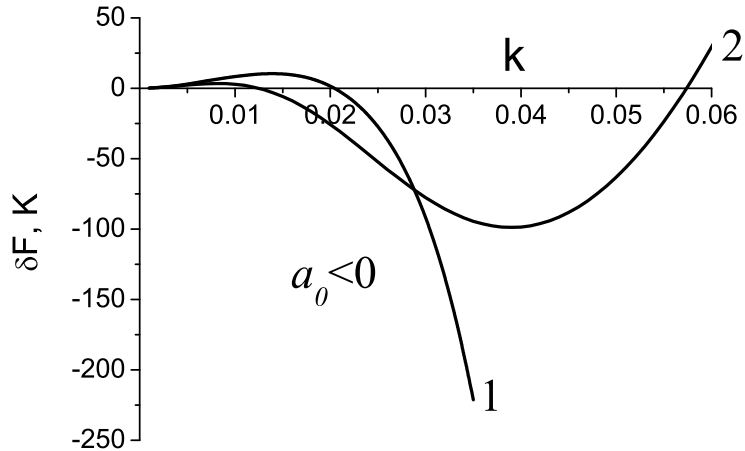}
\end{center}
\caption{Gain in the free energy due to electron localization in
the gas with negative scattering length as a function of the
variational parameter $k$ in linear approximation (curve 1) and
taking into account the non-linear behaviour of
 $V_0(n)$ (curve 2).}
\label{Fig_Negative_a0}
\end{figure}
\begin{figure}
\begin{center}
\includegraphics*[width=11cm]{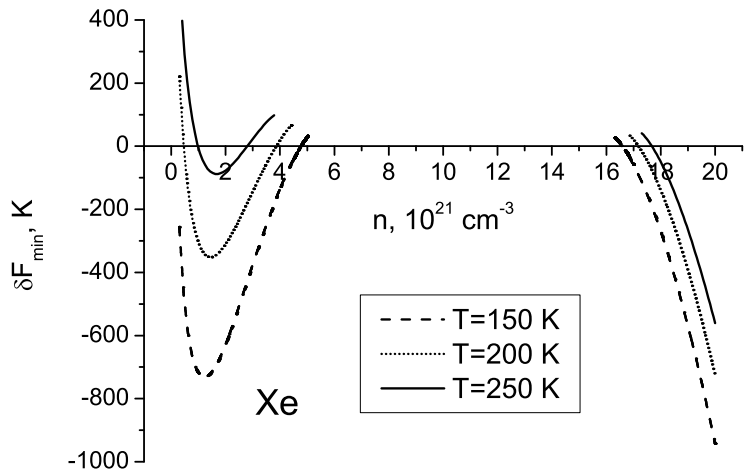}
\end{center}
\caption{Gain in free energy due to electron localization in Xe as
a function of gas atom number density at three different
temperatures.} \label{Fig_Xe_F_min}
\end{figure}
\begin{figure}
\begin{center}
\includegraphics*[width=11cm]{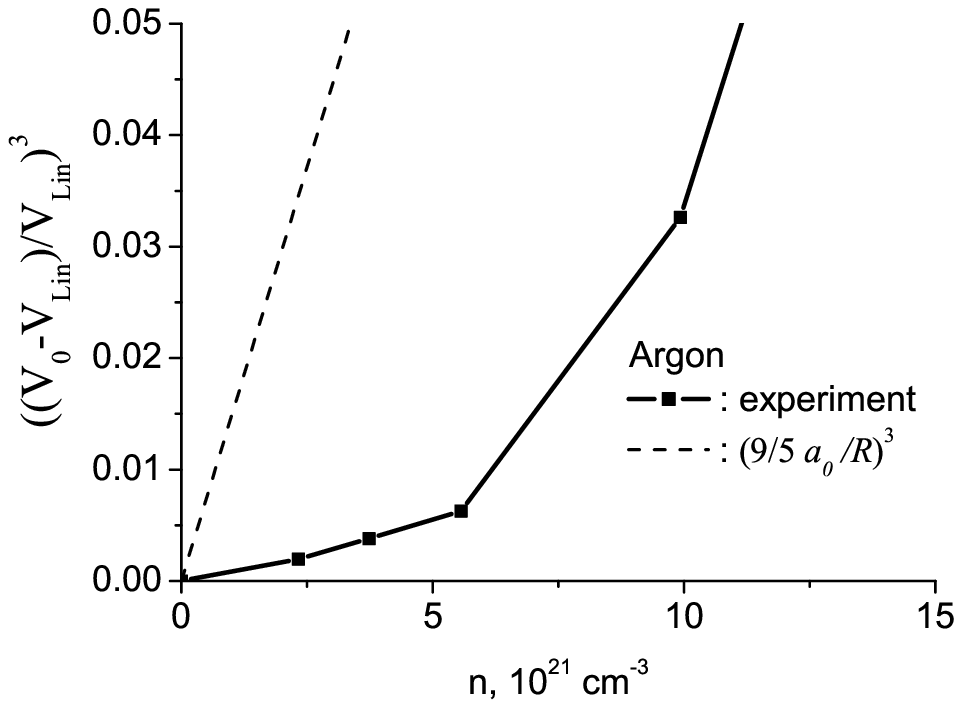}
\end{center}
\caption{Deviations from linear behaviour of $V_0(n)$ at small $n$
for Ar. $V_{\rm Lin} = \frac{2\pi\hbar^2a_0}{m}n$.}
\label{Fig_Ar_Low_n}
\end{figure}

Formally, the problem of finding the ground state of a single
electron in the gaseous media reduces to the minimization free
energy $F$ of the total system electron+gas with respect to
variations of the (spherically symmetric) electron wave function
$\psi(r)$ and gas tom number density $n(r)$
\cite{LevineSanders,ShikinUFN},
    \be
         F = \int d^3r \tilde{F}(r),\quad
         \tilde{F} = \frac{|\nabla\psi|^2}{2m} + nT\ln(nB) +
         \tilde{F}_{int}, \quad
         \tilde{F}_{int} = V(n)\psi^2(r),
         \label{FreeEnergy}
    \ee
where $B(T)$ is a function of temperature which does not affect
final results and therefore is not specified here. This procedure
results in a set of two coupled equations for $n(r)$ and
$\psi(r)$:
    \be
        -\frac{\hbar^2}{2m}\nabla^2\psi+V(r)\psi=E\psi,\quad V(r)=V_0(n(r))
        \label{ShroedEqu}
    \ee
    \be
        4\pi\int_0^{\infty}\psi^2(r)r^2dr=1, \nonumber
    \ee
    \be
        n(r)=n_g\exp\left[-\frac{|\psi|^2\partial V_0/\partial
        n}{T}\right],
        \label{BoltzmannEqu}
    \ee
where $n_g$ is the gas atom number density at infinity, $n(r)$ is
the local atom number density, $\psi(r)$ is the electron wave
function normalized to unity, and $T$ is the temperature.

To solve the equation set (\ref{ShroedEqu})-(\ref{BoltzmannEqu})
we employed (just as in Refs. \cite{LevineSanders,ShikinUFN}) the
variational approach with $\psi(r)$ selected in the form
    \be
        \psi(r,k) = \left(\frac{2}{\pi}\right)^{3/2}e^{-kr}.
        \label{VarFun}
    \ee
Here the variational parameter $k$ measures the electron
localization. By substituting Eq. (\ref{VarFun}) into Eq.
(\ref{BoltzmannEqu}) and finding $n(r,k)$ one can calculate the
free energy of the system $F(k)$ (\ref{FreeEnergy}). To study the
possibility of electron autolocalization at given $n_g$ and $T$
one should then plot the curve $\delta F(k)=F(k)-F_{deloc}$ (where
$\displaystyle F_{deloc} = \frac{2\pi \hbar^2 a_0}{m}n +
NT\ln(nB(T))$ is the total free energy of system consisting of a
uniform gas and delocalized electron described by the wave
function $\psi(r) = \mbox{\rm const}$) and check if this curve has
a minimum which is sufficiently deep compared to the temperature.
For single electron bubbles, where $a_0>0$, this program was
realized in \cite{LevineSanders,ShikinUFN} where the linaer
approximation for $V_0$ was employed. The single electron bubble
formation proves energetically favourable at sufficiently low
temperatures and sufficiently high densities (threshold values of
temperature and density follow the relation $T\sim n^{2/3}$), and
all the parameters of arising bubble well satisfy the adopted
assumptions: the bubble size is much larger than the interatomic
distance, the free energy minimum depth substantially exceeds
temperature, etc. We omit any quantitative details since for
$a_0>0$ the non-linearity of $V_0(n)$ does not introduce any
qualitative corrections to the bubble parameters and the resulting
picture is practically identical to that obtained earlier
\cite{LevineSanders,ShikinUFN}.

In the problem with $a_0<0$ we first mention that the relation
    \be
        n_0(r,k)=n_g\exp{[+\frac{2\pi\hbar^2|a_0||\psi(r,k)|^2}{mT}]}
        \label{BoltzmannLinear}
    \ee
following from Eqs. (\ref{BoltzmannEqu}) and (\ref{OpticalApprox})
leads in the linear theory to an unavoidable singularity in the
density distribution $n_0(r\to 0)$ (nothing can prevent the
arbitrary strong shrinking of electron wave function and the
corresponding growth of the gas density at the center of the
cluster resulting in infinite reduction of the system free energy)
as illustrated by variational calculations which yield for $\delta
F(k)$ the results plotted in Fig. \ref{Fig_Negative_a0} (curve 1).

By employing a more general expression for $n(r,k)$ with $V_0(n))$
(\ref{VFit}) it is easy to see that the trend towards density
enhancement around the localized electron taking place at
relatively large distances from the cluster core and correctly
described by Eq. (\ref{BoltzmannLinear}) is stopped near the
cluster center where the derivative $\partial V_0/\partial n $
changes its sign. It is also qualitatively clear that the halt in
the density growth is actually important if the uniform gas
density $n_g$ far from the cluster core is sufficiently low,
$n_g<\tilde n_{max}\le n_{min}$. Here $\tilde n_{max}\sim n(r\to
0)$ is the maximal gas density in the cluster core and $n_{min}$
is illustrated in \ref{Fig_V_NonLin}. If the inequality $n_g \sim
n_{min}$ is satisfied, the cluster formation mechanism defined by
Eqs. (\ref{ShroedEqu},\ref{BoltzmannEqu}) becomes inefficient (no
energy gain can be acquired by tuning the gas density to its
optimal value in the vicinity of the cluster center), and that is
actually why electrons in heavy inert gases behave as practically
free particles for gas densities close to $n_{min}$.

Now that the singularity suppression mechanism is clear, one can
apply the outlined variational procedure to quantitatively test
the above qualitative picture concerning the possibility of
electron cluster formation. Calculations reveal at not too high
temperatures $T$ there do exist density ranges where the free
energy gain due to electron localization $\delta F(k)$ as a
function of $k$ has a minimum with depth exceeding $T$ (curve 2 in
Fig. \ref{Fig_Negative_a0}). Numerical results for Xe are shown in
Fig. \ref{Fig_Xe_F_min} where the free energy gain calculated for
electron wave function defined by Eq. (\ref{VarFun}) and optimized
with respect to $k$ is plotted. It is clearly seen that the
localized state is only energetically favourable for not too low
densities outside some interval around $n_{min}$; the
characteristic cluster radii prove to be 10--20 $a_B$.

Hence, the non-linear corrections to the interaction energy
(\ref{OpticalApprox}) behave in qualitatively different ways for
$a_0>0$ and $a_0<0$. For positive scattering lengths non-linear
corrections to Eq. (\ref{OpticalApprox}) only slightly modifies
the overall picture of electron localization arising in the linear
approach. On the contrary, for negative scattering lengths the
presence of non-linearity in $V_0(n)$ becomes critically important
since it is the only factor capable of preventing the cluster from
shrinking to the Bohr length scale. Therefore, it is very
desirable to study the deviation of $V_0(n)$ from the linear
approximation (\ref{OpticalApprox}) at least for small $n$.
However, in spite of the fact that the problem of calculating
$V_0(n)$ has been addressed in many works (e.g., see Refs.
\cite{JortnerIakubovPogosov}), currently available theoretical
results are mainly numerical in nature and derived by replacing
the disordered medium with imaginary crystalline solid consisting
of the gas atoms with appropriate density after which the
conduction band bottom is calculated within the Wigner-Seitz
model. Major efforts in these works have been concentrated on
choosing the optimal pseudopotential describing the free electron
interaction with the inert gas atom closed shell and correct
screening of the long-range attracting potential $-\alpha
e^2/2r^4$ due to the Coulomb interaction between the electron and
polarizable gas atom, $\alpha$ being the atom polarizability. On
the other hand, it is interesting to note that for short-ranged
potentials the Wigner-Seitz model allows finding the asymptotic
behaviour of $V_0(n)$ at small $n$ beyond the linear approximation
through the scattering length $a_0$. Indeed, in that case the
requirement of vanishing of the wave function first derivative at
the spherical cell boundary (whose radius tends to infinity as
$n\rightarrow 0$) results in the following expression for the
conduction band bottom:
    \be
         V_0(n)=\frac{2\pi \hbar^2 a_0 n}{m}\left(1+\frac{9}{5}\frac{a_0}{R} +
                                      O\left(\frac{a_0^2}{R^2}\right) \right).
         \label{V_9_5}
    \ee
where $R = \left(\frac{3}{4\pi n}\right)^{1/3}$. It is seen that
the relative corrections to the linear approximation
(\ref{OpticalApprox}) are proportional to the small parameter
$a_0/R \ll 1$ which is the prediction that can be tested
experimentally. As an example, plotted in Fig. \ref{Fig_Ar_Low_n}
are experimental data on low density behaviour of $V_0(n)$ for
argon in the $((V_0-V_{\rm Lin})/V_{\rm Lin})^3, n)$ coordinates,
where $V_{\rm Lin} = V_0$ from Eq. (\ref{OpticalApprox}). It is
obvious that $\delta V_0(n)/V_{\rm Lin} \propto a_0/R$, although
the experimental proportionality coefficient is different from
that predicted by Eq. (\ref{V_9_5}). The reason for this
discrepancy is most likely the long-ranged nature of the effective
potential for electron interaction with the gas atom containing
the polarization contribution obeying the $R^{-4}$ law.

Thus, by taking into account the non-linear behaviour of $V_0(n)$,
it is possible extend the existing theory of electron
autolocalization in dense gases with positive scattering lengths
(single-electron bubbles in helium) to electrons in inert gases
with negative scattering lengths and describe possible formation
of electron clusters in these media. The clusters can arise at gas
densities both lower and higher than $n_{min}$ and are not formed
at densities close to $n_{min}$. The outlined picture is
consistent with available data on electron mobility $\mu$ in dense
cryogenic gases. The point is that with growing $n$ the
possibility of interpreting the mobility $\mu$ in terms of
single-particle collisions between the electrons and gas atoms is
gradually lost. However, if under these conditions the electron
still remains in an almost free non-localized state as suggested
by the above analysis, it is natural to describe its interaction
energy with the gaseous media responsible for scattering by the
expression \cite{BasakCohen}
    \be
        \delta V_0=\frac{\partial V_0(n)}{\partial n}\delta n
    \ee
where $\delta n$ is the gas density fluctuation of the thermal
origin. It is then obvious that the derivative $\partial
V_0(n)/\partial n$ vanishing at $n=n_{min}$ yields a peak in the
density density dependence of electron mobility. Hence,
experimental observation of electron mobility peaks in all three
heavy inert gases around the respective values of $n_{min}$ can be
considered as a confirmation of the absence of electron
localization in the vicinity of the mobility peak.

This work was supported by Russian Foundation for Basic Research
grant \# 06-02-17121 and the Program ``Physics of Condensed
Matter'' of the Presidium of Russian Academy of Sciences.


\begin{thebibliography}{00}

\bibitem{HuangFreeman} S.Huang, G.Feeman J.Chem.Phys., 68 (1978) 1355

\bibitem{MillerHoveSpear} L.Miller, S.Hove, W.Spear. Phys. Rev., 166 (1968)
871.

\bibitem{Christophorou} L.Christophorou, R.Blaunstein, D.Pittman. Chem. Phys. Lett., 18
(1973) 509.

\bibitem{LastBook} Electronic Excitations in Liquefied Rare Gases. Eds. W.F.Schmidt and E.Illenberger.
American Scientific Publishers, Stevenson Ranch, USA, 2005

\bibitem{Khrapak_ZHETF} A.Khrapak, I.Jakubov.  ZhETF, 69 (1975)
2042. (in Russian)

\bibitem{Iakubov160} I.Jakubov, A.Khrapak.  Chem. Phys. Lett., 39 (1976)
160.

\bibitem{KhrapakBook} I.Jakubov, A.Khrapak. {\sl Electrons in Dense Gases and Plasma}. Moscow, Nauka,
1981. (in Russian)

\bibitem{Fermi1934} E.Fermi, Nuovo Cim., 11, (1934), 157

\bibitem{LifshitzSondheimer} I.M.Lifshits, S.A.Gredeskul, L.A.Pastur.
{\sl Introduction to the theory of Disordered Systems} Moscow, Nauka, 1982; S.Doniach and
E.H.Sondheimer. Green's Functions for Solid State Physicists.
Imperial College Press, 1998.

\bibitem{NonLin_Experimental} For example, see Chapter 3 in Ref. [4].

\bibitem{LevineSanders} J.L.Levine, T.M.Sanders. Phys. Rev., 154 (1967) 138.

\bibitem{ShikinUFN} V.B.Shikin. Uspekhi Fiz. Nauk, 121 (1976) 457.
(in Russian)

\bibitem{JortnerIakubovPogosov} B.Springett, M.Cohen, J.Jortner. Phys.Rev.159, (1967), 183;
I.Jakubov, V.Pogosov, Phys. Rev., B51 (1995) 14941; I.Jakubov,
V.Pogosov. Phys. Rev., B53 (1996) 13362.

\bibitem{BasakCohen} S.Basak, M.H.Cohen. Phys. Rev., B20 (1979) 3404.

\end{thebibliography}
\end{document}